\documentclass[twocolumn,showpacs,preprintnumbers]{revtex4}%
\usepackage{amsfonts}
\usepackage{pstricks}
\usepackage{pst-poly}
\usepackage{graphicx}
\usepackage{txfonts}
\usepackage{times}
\usepackage{multirow}
\usepackage{amssymb}
\usepackage{subeqnarray}

\def\bi{\begin{itemize}}
\def\ei{\end{itemize}}
\def\be{\begin{equation}}
\def\ee{\end{equation}}
\def\bea{\begin{eqnarray}}
\def\eea{\end{eqnarray}}
\def\ben{\begin{eqnarray*}}
\def\een{\end{eqnarray*}}
\def\non{\nonumber}

\begin{document}

\title{Multi-stability in an optomechanical system with two-component Bose-Einstein condensate}
\author{Ying Dong$^{1,2}$}
%\email{yingdong@rice.edu}
\author{Jinwu Ye$^{3,4}$}
%\email{}
\author{Han Pu$^1$}
%\email{hpu@rice.edu}
 \affiliation{$^1$Department of
Physics and Astronomy, and Rice Quantum Institute, Rice University, Houston, Texas 77251-1892, USA\\
$^2$Hefei National Laboratory
for Physical Sciences at Microscale and Department of Modern
Physics, University of Science
and Technology of China, Hefei 230026, P. R. China \\
$^3$Department of Physics, Capital Normal University, Beijing 100048, P. R. China\\
$^4$Department of Physics, The Pennsylvania State University, University Park, PA 16802, USA }
\date{\today}

\begin{abstract}
We investigate a system consisting of a two-component Bose-Einstein condensate interacting dispersively with a Fabry-Perot optical cavity where the two components of the condensate are resonantly coupled to each other by another classical field. The key feature of this system is that the atomic motional degrees of freedom and the internal pseudo-spin degrees of freedom are coupled to the cavity field simultaneously, hence an effective spin-orbital coupling within the condensate is induced by the cavity. The interplay among the atomic center-of-mass motion, the atomic collective spin and the cavity field leads to a strong nonlinearity, resulting in multi-stable behavior in both matter wave and light wave at the few-photon level.
\end{abstract}

\pacs{03.75.Mn, 03.75.Kk, 42.50.Pq, 42.65.-k}

\maketitle

%%%%%Introduction%%%%%Introduction%%%%%Introduction%%%%%Introduction

In the past few years, there has been a great surge of interest in the nonlinear phenomena associated with the so called optomechanical systems, which are realized by coupling a mechanical oscillator to an electromagnetic field in a cavity \cite{opto1,opto2,opto3,opto4,opto5,opto6,opto7}.
The exploration of these systems has led to many exciting developments, including self-sustained oscillations \cite{self1,self2-bista}, bistability \cite{self2-bista,bista}, and optomechanical chaos \cite{Chaos}. Besides these linear optomechanical coupling systems, devices with nonlinear optomechanical coupling have been studied in some very recent experimental works \cite{non-cou1,non-cou2} as well.

On the other hand, more recent studies on the cavity quantum electrodynamics (QED) with an ensemble of ultracold atoms, both bosonic \cite{bose1,bose2,bose3,bose4,bose5} and fermionic \cite{fermi}, give rise to a new platform of cavity optomechanics. In this new regime of cavity QED, a cavity field  at the level of few or even a single photon can significantly affect the collective motion of the whole atomic samples. This allows us to study the nonlinear dynamics of ultracold atomic gases in a new domain.

Among all these nonlinear phenomenon in a cavity optomechanical system, the bistable behavior is one of the focuses of research interest. The optical bistability in optomechanical systems has been studied both in theory \cite{bista-theo} and in experiment \cite{bista-atom1,bista-atom2}. Strong matter wave bistability has also been investigated in a spinor Bose-Einstein condensate (BEC)\cite{bista-matter}.
In this work, we propose a scheme to exploit the multistable behavior in a two-component BEC coupled to a Fabry-Perot cavity. Here the two components of the condensate are coupled by another classical optical field, hence realizing a pseudo-spin half system. The cavity supports a single-mode standing wave optical field, which interacts with atoms dispersively. When the coupling strengths of the two atomic spin components and the optical field are different, the cavity field will then couple to both the external center-of-mass and the internal spin degrees of freedom of the condensate. Whereas in previous studies of BEC-cavity system, the cavity field couples either to the center-of-mass \cite{bose1,bose2,bose3,bose4,bista-theo,bista-atom1,bista-atom2} or to the spin degrees of freedom \cite{bista-matter}, but not to both \cite{larson}. As we shall demonstrate, the nonlinear coupling among the external and internal states of the condensate and the cavity photons leads to multi-stability in both light wave and matter wave.

%%%%%model%%%%%model%%%%%model%%%%%model%%%%%model%%%%%model%%%%%model
\begin{figure}
\includegraphics[width=3.3in]{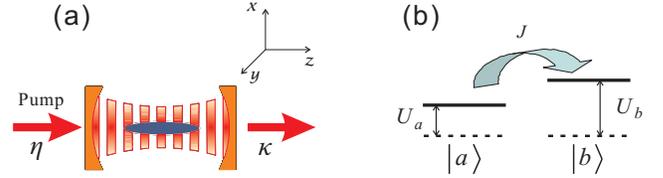}
\caption{ (Color Online) (a) Schematic diagram of the system. The cavity is pumped at rate $\eta$ and the cavity photon decays at rate $\kappa$. (b) Atomic level diagram: The two stable atomic hyperfine ground states $|a\rangle$ and $|b\rangle$ are coupled by a classical field with coupling strength $J$. In addition, they interact with the cavity dispersively, resulting to an energy shift $U_a$ and $U_b$, respectively.}
\end{figure}

In our model, as depicted schematically in Fig. 1(a), we consider a quasi-one-dimensional BEC of $N$ atoms in two stable hyperfine spin ground states $|a\rangle$ and $|b\rangle$ trapped in an high-finesse Fabry-Perot cavity. The two spin states are coupled to each other by a classical light field (This can be, for example, either a two-photon Raman field or an Radio-Frequency field).
The BEC and the cavity are in the strong coupling regime of cavity QED, that is, the maximum coupling strength between a single atom and a single intra-cavity photon, $g_a$ and $g_b$, are larger than both the amplitude decay rate of the intra-cavity field $\kappa$ and that of the atomic excited state. The coupled dynamics of the BEC and the cavity field is driven by continuously applying a weak pump laser field with frequency $\omega_p$ and amplitude $\eta$ along the cavity axis, taken to be the $z$-axis.

Let $c$ ($c^\dag$) denote the annihilation (creation) operators for the single-mode cavity field, and $a$ ($a^{\dag}$) and $b$ ($b^\dag$) the annihilation (creation) field operators for the atoms in state $|a\rangle$ and $|b \rangle$, respectively.
Then the Hamiltonian in a frame rotating at the pump frequency $\omega_p$ can be written as (to focus on the atom-photon interaction, we have neglected the atom-atom collisions)
\begin{eqnarray}\label{ham}
H&=&\int_0^L\mathrm{d} z \,\left[a^{\dagger}(z)\frac{-\hbar^2}{2m} \frac{\partial^2}{\partial z^2}a(z)+
b^{\dagger}(z)\frac{-\hbar^2}{2m} \frac{\partial^2}{\partial z^2} b(z)\right. \nonumber\\
&&+\hbar Ja^{\dagger}(z)b(z)+\hbar J b^{\dagger}(z)a(z) \nonumber\\
&&+\hbar\big(U_aa^{\dag}(z)a(z)+U_bb^{\dag}(z)b(z)\big) \,\mathrm{cos}^2(kz)c^{\dag}c\bigg]\nonumber\\
&&-\hbar \delta_cc^{\dag}c+i\hbar\eta(c^{\dag}-c)\,.
\end{eqnarray}
Here $L$ is the length of the cavity and $\delta_c=\omega_p-\omega_c$ is the cavity-pump detuning. The second line of Eq.~(\ref{ham}) represents the coupling between the two spin states by the classical field, with $J$ being the coupling strength (without loss of generality, we take $J$ to be real and negative). $U_i=g_i^2/(\omega_p-\omega_i)\quad(i=a,b)$ characterizes the cavity field-induced energy shift of the atomic spin states, with $g_i$ being the resonant coupling strength between the atom and the cavity field, and $\omega_i$ the transition frequency for the two atomic states, respectively. We assume that the pump-atom detuning is large enough so that the atomic upper level can be adiabatically eliminated and the interaction between the cavity photon and the atom is essentially of dispersive nature.

The standing-wave cavity mode couples different momentum modes of the condensate separated by an integer multiple of $2\hbar k$, with $k$ being the wavenumber of the cavity photon. Recent experiments \cite{bista-atom1} suggest that, when cavity photon number is not large, the atomic momentum modes interacting significantly with the cavity field are those with momenta 0 and $\pm 2\hbar k$. Neglecting all higher order modes, we can expand the atomic field operators in $k$-space as $a(z)=\left(a_0+\sqrt{2}a_2\cos 2kz\right)/\sqrt{L}$ and $b(z)=\left(b_0+\sqrt{2}b_2\cos 2kz\right)/\sqrt{L}$. Substituting them into Eq.~(\ref{ham}) leads to the Hamiltonian under this few-mode approximation:
\begin{eqnarray}\label{ham2}
\frac{H}{\hbar}&=&\omega(a^{\dag}_2a_2+b^{\dag}_2b_2)+J(a_0^{\dag}b_0+a_2^{\dag}b_2)+J(b_0^{\dag}a_0+b_2^{\dag}a_2)\nonumber\\
&&+\frac{1}{2}\bigg[U_a(a^{\dag}_0a_0+a^{\dag}_2a_2+\frac{1}{\sqrt{2}}a^{\dag}_0a_2+\frac{1}{\sqrt{2}}a^{\dag}_2a_0)\nonumber\\ &&+U_b(b^{\dag}_0b_0+b^{\dag}_2b_2+\frac{1}{\sqrt{2}}b^{\dag}_0b_2+\frac{1}{\sqrt{2}}b^{\dag}_2b_0)\nonumber\\ &&-2\delta_c\bigg]c^{\dag}c+i\eta \,(c^{\dag}-c)\,,
\end{eqnarray}
where $\omega=4\hbar k^2/2m$ is the photon recoil frequency.

Now we can draw an analogy with the optomechanical system by defining two harmonic oscillator modes with displacement $\hat{X}_a=(a^{\dag}_0a_2+a^{\dag}_2a_0)/\sqrt{2N_a}$ and $\hat{X}_b=(b^{\dag}_0b_2+b^{\dag}_2b_0)/\sqrt{2N_b}$. Their corresponding conjugate variables are $\hat{P}_a=-i(a^{\dag}_0a_2-a^{\dag}_2a_0)/\sqrt{2N_a}$ and $\hat{P}_b=-i(b^{\dag}_0b_2-b^{\dag}_2b_0)/\sqrt{2N_b}$, respectively. $N_s$ denotes the number of atoms in spin state $|s=a,b\rangle$. Under the condition that most of the atoms are still in the zero-momentum mode, which is an excellent approximation for the parameters of current experiments \cite{bista-atom1}, it is straightforward to verify that these variables satisfy the commutation relations $[\hat{X}_j,\hat{P}_j]=i(N_{j_0}-N_{j_2})/N_j \simeq i\quad(j=a,b)$, and all other commutaters vanish.

In the following we adopt a mean-field treatment by replacing the operators $a_0$($b_0$) and $a_2$($b_2$) with their corresponding $\mathcal{C}$ numbers $\alpha_0=\sqrt{N_{a_0}}e^{-i\theta_{a_0}}$ ($\beta_0=\sqrt{N_{b_0}}e^{-i\theta_{b_0}}$) and $\alpha_2=\sqrt{N_{a_2}}e^{-i\theta_{a_2}}$ ($\beta_2=\sqrt{N_{b_2}}e^{-i\theta_{b_2}}$).
Now we can define another pair of conjugate variables as the atomic collective spin $\hat{M}=\hat{N}_a-\hat{N}_b$ and the relative phase $\theta=\theta_{a_0}-\theta_{b_0}$. Obviously we have $[\hat{X}_i,\hat{M}]=[\hat{P}_i,\hat{M}]=0$ and $[\hat{X}_i,\theta]=[\hat{P}_i,\theta]=0\quad(i=a,b)$. In other words, we have three pairs of independent conjugate variables: $(\hat{X}_a,\hat{P}_a)$, $(\hat{X}_b,\hat{P}_b)$ and $(\hat{M}, \theta)$. The first two pairs represent the oscillator modes and originate from the center-of-mass motion of the condensate, while the last pair originate from the internal motion of the condensate.
We can then rewrite the Hamiltonian under the few-mode approximation in terms of these three pairs of conjugate variables as
\begin{eqnarray}\label{ham3}
\frac{H}{\hbar}&=&\frac{\omega}{2}(\hat{X}_a^2+\hat{P}_a^2+\hat{X}_b^2+\hat{P}_b^2)+J\sqrt{N^2-\hat{M}^2}\cos\theta\non\\
&&+\frac{1}{2} \left[(U_a-U_b)\hat{M}/2+U_a\hat{X}_a\sqrt{(N+\hat{M})/2} \right. \nonumber \\
&&+\left. U_b\hat{X}_b\sqrt{(N-\hat{M})/2}-\Delta_c \right]c^{\dag}c+i\eta \,(c^{\dag}-c)\,,
\end{eqnarray}
where $\Delta_c=2\delta_c-(U_a+U_b)N/2$.

To proceed further, we treat the leakage of cavity photons phenomenologically by introducing a cavity decay rate $\kappa$ whose typical value ($\sim$ 1 MHz) is much larger than $\omega$ and $J$, under which condition we can assume that the cavity field always follows adiabatically the atomic dynamics. From $i\hbar\dot{c}=[c,H]=0$, we obtain the mean intracavity photon number $N_c$ as
\begin{equation}\label{photo}
\langle c^{\dag}c\rangle=\frac{\bar{\eta}^2}{1+\frac{1}{4}\left[\bar{U}_mm+\bar{U}_ax_a\sqrt{\frac{1+m}{2}}
+\bar{U}_bx_b\sqrt{\frac{1-m}{2}}-\bar{\Delta}_c \right]^2}\,.
\end{equation}
Here we have adopted the normalized variables $x_a=X_a/\sqrt{N}$, $x_b=X_b/\sqrt{N}$, $m=M/N$ and the dimensionless parameters $\bar{\eta}=\eta/\kappa$, $\bar{U}_a=U_aN/\kappa$, $\bar{U}_b=U_bN/\kappa$, $\bar{U}_m=(\bar{U}_a-\bar{U}_b)/2$, $\bar{\Delta}_c=\Delta_c/\kappa$. For convenience of the following discussion, we also define two other dimensionless parameters as
$\bar{\omega}=\omega N/\kappa$, $\bar{J}=JN/\kappa$.

We can now write down the equations of motion as follow:
\begin{subeqnarray}
\dot{x}_a&=&\bar{\omega}p_a \,, \\
\dot{x}_b&=&\bar{\omega}p_b \,,\\
\dot{m}&=&2\bar{J}\sqrt{1-m^2}\sin\theta \,, \label{theta1} \\
\dot{p}_a&=&-\bar{\omega}x_a-\bar{U}_a\sqrt{(1+m)/8} \, \langle c^{\dag}c\rangle \,,\\
\dot{p}_b&=&-\bar{\omega}x_b-\bar{U}_b\sqrt{(1-m)/8} \, \langle c^{\dag}c\rangle \,,\\
\dot{\theta}&=&2\bar{J}m\cos\theta/\sqrt{1-m^2}-\left[ \bar{U}_m+\bar{U}_ax_a/\sqrt{8(1+m)} \right.  \nonumber  \\
&&-\left. \bar{U}_bx_b/\sqrt{8(1-m)} \right]\,\langle c^{\dag}c\rangle \,,
\label{motion}
\end{subeqnarray}
where the time derivatives are taken with respect to the dimensionless time $\tau=\kappa t$.

Combining Eqs.~(\ref{photo}) and (\ref{motion}), one can find the effective Hamiltonian $H_{\rm eff}$ as
\begin{eqnarray}\label{ham4}
\frac{H_{\rm eff}}{\hbar\kappa}&=&\frac{\bar{\omega}}{2}({x}_a^2+{p}_a^2+{x}_b^2+{p}_b^2)+\bar{J}\sqrt{1-{m}^2}\cos\theta\non\\
&&-\bar{\eta}^2\arctan\bigg[\frac{1}{2} \left(\bar{\Delta}_c-\bar{U}_mm-\bar{U}_ax_a\sqrt{\frac{1+m}{2}} \right. \non\\
&& \left. -\bar{U}_bx_b\sqrt{\frac{1-m}{2}} \right)\bigg]\,. \label{heff}
\end{eqnarray}

\begin{figure}
\includegraphics[scale=0.8]{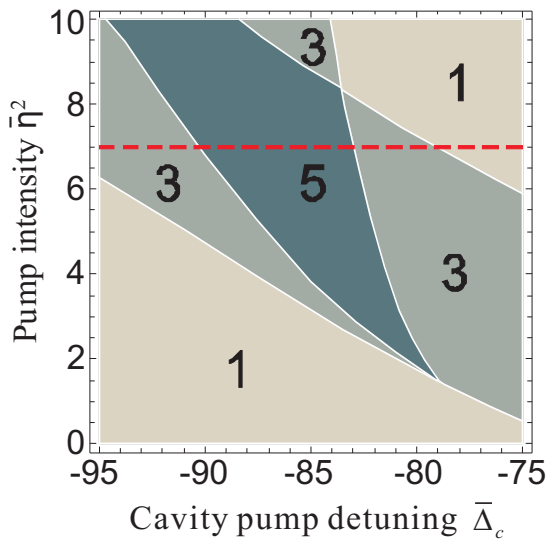}
\caption{(Color Online) Phase diagram for different types of solutions with $\theta=0$ in the parameter space of $\bar{\eta}^2$ and $\bar{\Delta}_c$. Different regions are differentiated by their colors. The digits labeled in each region denote the numbers of corresponding solutions. The dimensionless parameters are taken to be $\bar{U}_a=50$, $\bar{U}_b=200$, $\bar{\omega}=400$, $\bar{J}=-1$. The total number of the atoms are set as $N=10^5$. The red dashed line corresponds to $\bar{\eta}^2=7$.}
\label{number}
\end{figure}

From Eqs.~(\ref{motion}) one can see that the atomic dynamics will depend on the intra-cavity photon number. Conversely, the intracavity photon number is dependent upon the atomic variables via Eq.~(\ref{photo}). It is this interdependence of the atomic and photonic modes that leads to interesting multistability of this coupled system, which will be the main result of this work.

The steady state solutions are obtained by setting the time derivatives to zero in Eqs.~(\ref{motion}) which yields six coupled nonlinear algebraic equations. In the case of $U_a=U_b$, i.e., when the light shifts induced by the cavity photon are spin-independent, it is not difficult to see that $m=0$ in the steady state. In other words, under this situation, the populations in the two spin states are always equal to each other, which effectively freezes the atomic spin degrees of freedom. Thus, we will always focus on the case where $U_a \neq U_b$.

We may get some useful information from the phase diagram identifying different types of phase-dependent solutions. It follows from Eq.~(\ref{theta1}) that $\theta=0$ or $\pi$ in the steady state. We will focus only on the $\theta=0$ branch, which for $J<0$ represents the lower energy branch. In the parameter space of $\bar{\eta}^2$ and $\bar{\Delta}_c$, the number of steady-state solution are illustrated in Fig.~\ref{number}. We can see that, in certain parameter regimes, the number of different solutions of the system can be more than one, which indicates that multi-stable behavior may be observed. By varying $\bar{\eta}$ and/or $\bar{\Delta}_c$, one can traverse different solution regions of the system. So, this coupled system can be easily manipulated by tuning the intensity or frequency of the pump laser field.

\begin{figure}
\includegraphics[width=2.8in]{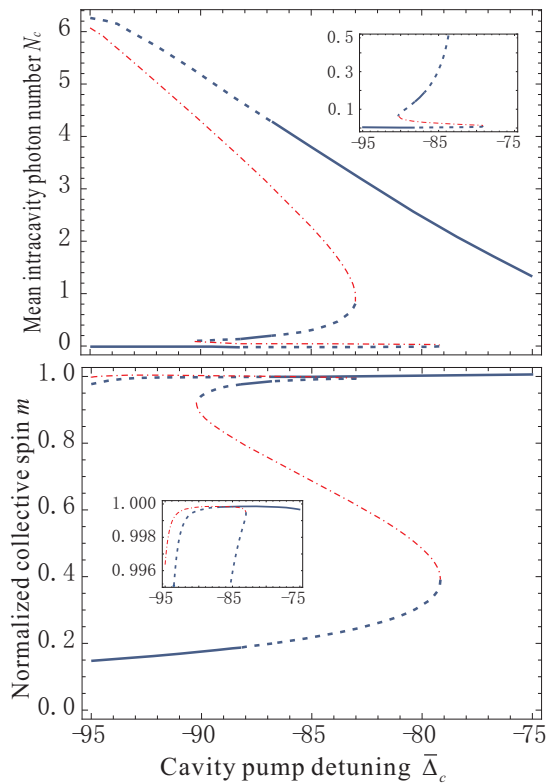}
\caption{(Color Online) Mean intracavity photon number $N_c$ (upper panel) and normalized collective spin $m$ (lower panel) versus cavity pump detuning $\bar{\Delta}_c$ for the steady-state solutions with $\bar{\eta}^2=7$, corresponding to the red-dashed line in Fig. 2. The branches represented by the blue (darker) lines correspond to the dynamically stable solutions (the solid parts denote the ground state and the dashed parts denote stable but not the ground state), and the ones represent by the red (lighter) dot-dashed lines correspond to the dynamically unstable solutions.}
\label{solution}
\end{figure}

As an example, we consider the case where the pump intensity is fixed at $\bar{\eta}^2=7$. By varying the cavity-pump detuning $\bar{\Delta}_c$, the equilibrium properties of the system are changed, as shown by the red-dashed line in Fig.~\ref{number}. The corresponding solutions are derived and the typical results are shown in Fig.~\ref{solution}. From these plots, one can see that, both the cavity field and the atomic spin population exhibit multistable behavior. For certain values of detuning $\bar{\Delta}_c$, it supports three or five steady-state solutions. A standard linear stability analysis shows that in the region with three solutions, two of them are dynamically stable and in the region with five solutions, three of them are dynamically stable. Hence these represent bi- and tri-stable regimes, respectively. In these multi-stable regimes, we calculate the energies of the stable states according to Eq.~(\ref{heff}), from which we identify the ground-state solution which are represented by the solid lines in Fig.~\ref{solution}. As can be seen, the ground state jumps from one branch to another at certain critical values as $\bar{\Delta}_c$ is scanned. These critical points correspond therefore to first-order transitions in this system.

We remark that optical tristable behavior has been discussed in theory several decades ago \cite{opto-tri1} and have been observed in various systems \cite{opto-tri2,opto-tri3,opto-tri4,opto-tri5}. Here we report a new platform where simultaneous bi- or tri-stability in optical and matter waves can be observed. We emphasize that, in the regime of weak cavity field such that the few-mode approximation is valid, the existence of tristable regime requires both external center-of-mass and the internal spin degrees of freedom of the condensate to be present. By contrast, in previous studies where one or the other of these degrees of freedom is absent, only bistable behavior is observed \cite{bose1,bose2,bose3,bose4,fermi,bista-theo,bista-atom1,bista-atom2,bista-matter}.

Note also that the presence of several degrees of freedom does not guarantee tristability. For example, optomechanical systems with multiple degrees of freedom have been studied in several works recently \cite{approx,multi1,multi2}, with no tristability being reported. There is a common feature in the systems studied in these works, that is each mechanical degrees of freedom is only coupled to the cavity mode, and there is no direct coupling among themselves. In fact, this corresponds to setting $J=0$ in our model, so that the collective spin $M$ becomes a conserved quantity and hence effectively freezes the spin degrees of freedom. Then, in principle, through a canonical transformation, one can always rewrite the Hamiltonian (see Eq.~(\ref{ham3}) for example) as a sum of several independent subsystems, each of which can be regarded as an oscillator coupled with a cavity field or just a free oscillator. Obviously, there would be no multi-stable behavior other than bistability in such a system.

In contrast, in the model we have considered here, the mechanical degrees of freedom represented by $X_a$ and $X_b$, the spin degrees of freedom represented by $M$ and the cavity field are all coupled simultaneously, as can be seen from Eq.~(\ref{ham3}). It is this nonlinear coupling that makes tristability possible.

In summary, we have studied the interaction of a two-component BEC with a standing-wave cavity field, where the two components are coupled by another classical optical field. We show that this coupled cavity-BEC system can display simultaneously optical multi-stability at the few-photon level and matter-wave multi-stability involving a whole condensate with a macroscopic number of atoms. This highly controllable optical and collective spin multi-stability can be very useful both in exploring fundamental quantum physics such as understanding decoherence in a macroscopic system and in applications such as building switches and logical gates for quantum information processing. Our study also opens up possibilities to explore nonlinear dynamical effects such as chaos \cite{chaos2} and bifurcation \cite{bifur} in an optomechanical system at the regime of few-photon level. In this work, we have adopted a few-mode approximation. As a self-consistent check, we have verified that for the parameters we used, the zero-momentum atomic population always exceeds $90\%$, which should make the approximation valid. In the future, it will be instructive to numerically study the validity regime of the few-mode approximation and explore the potentially interesting physics beyond the approximation. Another interesting revenue of research is to explore the novel physics induced by the coupling between the motional and spin degrees of freedom within the condensate.

%\acknowledgments
This work is supported by the NSF, the Welch Foundation (Grant No.
C-1669) and by a grant from the Army Research Office with funding
from the DARPA OLE Program. J. Ye is supported by NSF DMR-0966413.

\end{document}